\documentclass[prd, amsfonts, twocolumn, nofootinbib, showpacs]{revtex4}
\usepackage{graphicx, epsfig}
\usepackage{color}
\usepackage{amsmath}

\usepackage{tikz}
\usetikzlibrary{decorations.markings}
\def\({\left(}
\def\){\right)}
\def\[{\left[}
\def\]{\right]}

\def\f#1#2{\frac{#1}{#2}}
\def\g{\gamma}
\def\d{\partial}
\def\de{\delta}
\def\De{\Delta}

\def\e{\eta}

\def\k{\kappa}

\def\m{\mu}
\def\n{\nu}
\def\o{\omega}

\def\p{\pi}
\def\r{\rho}
\def\s{\sigma}

\def\th{\theta}

\def\<{\langle}
\def\>{\rangle}

\providecommand{\abs}[1]{\lvert#1\rvert}

\definecolor{orange}{rgb}{1,0.5,0}
\definecolor{test}{rgb}{.5,0.5,.5}

\preprint{APS/123-QED}
\begin{document}

\title{Brane Localization and Stabilization via Regional Physics}

\author{David M. Jacobs}%\footnote{Email address: david.m.jacobs@case.edu}}
\author{Glenn D. Starkman}%\footnote{Email address: glenn.starkman@case.edu}}
\author{Andrew J. Tolley}%\footnote{Email address: andrew.j.tolley@case.edu}}

\affiliation{Center for Education and Research in Cosmology and Astrophysics}
\affiliation{Department of Physics and Institute for the Science of Origins, \\Case Western Reserve University}

\begin{abstract}
%\tc{red}{\bf I'm in the middle of re-editting this PRL.  I'm made a backup of the older one. I felt that the whole discussion of local/regional/global physics was wonky and unclear. We haven't precisely defined what regionalism is nor had we really explain enough about it to warrant having it in the title. The whole point can be thought of as what might look like global effects are actually regional when one realizes that the manifold is an approximation to a region of manfold.}\\
Extra-dimensional scenarios have become widespread among particle and gravitational theories of physics to address several outstanding problems, including cosmic acceleration, the weak hierarchy problem, and the quantization of gravity.  In general, the topology and geometry of the full spacetime manifold will be non-trivial, even if our ordinary dimensions have the topology of their covering space.  Most compact manifolds are inhomogeneous, even if they admit a homogeneous geometry, and it will be physically relevant where in the extra-dimensions one is located. 
In this letter, we explore the use of both local and global effects in a braneworld scenario to naturally provide position-dependent forces that determine and stabilize the location of a single brane.  For illustrative purposes,  we consider the 2-dimensional hyperbolic horn and the Euclidean cone as toy models of the extra-dimensional manifold,
and add a brane wrapped around one of the two spatial dimensions. % (a 1-brane for $(2+1)$-dimensional space-time, a 4-brane for $(5+1)$d).
We calculate the total energy due to brane tension and bending (extrinsic curvature) as well as that due to the Casimir energy of a bulk scalar satisfying a Dirchlet boundary condition on the brane.  From the competition of at least two of these effects there can exist a stable minimum of the effective potential for the brane location.  However, on more generic spaces
(on which more symmetries are broken) any one of these effects may be sufficient to stabilize the brane. We discuss this as an example of physics that is neither local nor global, but {\it regional}.

%{\bf (Dave:  This might not be true. We should look at Ponton and Poppitz, 2001. "Casimir Energy and Radius Stabilization in Five and Six Dimensional Orbifolds" )}

\end{abstract}

\maketitle

%\sec{Introduction}

Extra spatial dimensions are a common feature of modern fundamental theories of physics beyond the Standard Model.  
They arise in attempts to unify the gauge theories of the Standard Model with gravity (e.g. string theory);
in solving the weak hierarchy problem with large extra dimensions (e.g. \cite{ArkaniHamed:1998rs,Antoniadis:1998ig,Randall:1999ee}); in explaining flavor hierarchy (e.g. \cite{ArkaniHamed:1999dc}); and in addressing the dark energy problem via infra-red modifications of gravity (e.g. \cite{Dvali:2000hr, Dvali:2007kt,deRham:2007xp}).

Two popular (and nonexclusive) ideas to make these scenarios consistent with experimental constraints are to confine Standard Model fields to submanifolds of lower dimension (branes) and also to compactify extra dimensions. In the latter case, a topology must be specified in order to determine the phenomenology; since topology is a \emph{global} property it cannot be determined by the \emph{local} Einstein equations.  

Manifolds of non-trivial topology are most familiarly obtained from a global covering space of constant curvature, then modding out by a discrete subgroup of the covering space's isometry group.  Thus a cylinder ($E^1 \times S^1$) is obtained from the Euclidean space (${E}^2$) by modding out by the group ($\Gamma$) of 1-dimensional discrete translations. The local geometry remains homogeneous and isotropic, however the behavior of fields indicates that some symmetry has been broken.  For example, momentum is quantized in the compact dimension, while it remains continuous in the infinite dimension; in other words the physics is locally anisotropic. Even so, \emph{there exist no special places in this space}.

Contrast this with the 2D horn, obtained from the hyperbolic space ${\cal H}^{2}$ and again modding out by $\Gamma$.  Fields on this space indicate that both rotational \emph{and} translational symmetry are broken. For example, modes of a given momentum are typically highly suppressed in the region where the wavelength is larger than the circumference of the compact dimension. This tells us e.g. the probability amplitude for local interaction between a brane embedded in this space and fields propagating in the bulk (e.g. the graviton) would be very sensitive to the brane position. Even the effective dimensionality of the bulk space is sensitive to the location of such a brane, as the energy required to excite modes in the compact dimension increases significantly as the brane moves down into the cusp.  %Since translation invariance is broken there exists a notion of absolute position.

A final example, and the one explored in detail in this work, is that the effective energy density associated with a 4-brane wrapped around this space depends on the location along the horn.   Different contributions to this energy -- brane tension, extrinsic curvature, the Casimir energy of bulk fields with non-trivial boundary conditions on the brane --  each vary with position, providing local forces on the brane. Thus we see that even on spacetimes which are locally homogeneous and isotropic the local physics will generally not be due to the global structure of the manifold.

Considering, however, that hyperbolic horns (which are not fully compact) possess some of the same distinguishing features as the ``cuspy" regions of compact hyperbolic manifolds (CHMs)\footnote{CHMs provide an appealing geometric solution for the hierarchy problem (see \cite{Kaloper:2000jb}) and for some cosmological problems \cite{Starkman:2000dy,Starkman:2001xu}; they could be considered a hyperbolic, $d>1$, version of models discussed in \cite{ArkaniHamed:1998rs} or \cite{Randall:1999ee}, wherein all but one modulus is fixed.}, the dependence of the aforementioned local effects on the full structure of the space is likely an overstatement.  What appeared to be a result of the global properties of the horn should just be attributed to the features of a finite \emph{region} of the actual manifold, which might be a CHM, for example.

%For this reason, the aforementioned physical effects may actually be \emph{regional} in nature, as opposed to global. 
% -- it may look that way because one is far enough down the narrow part of a CHM. 

Though presently lacking a precise mathematical definition, regional manifold properties are those which cannot be deduced from the local geometry alone and yet do not generally depend on the full structure of the manifold. An example regional quantity is readily found on the horn -- the shortest non-trivial closed curved through a given point, i.e. the circumference.

%This illustrates how in addition to the local and global properties of a manifold, some properties depend on intermediate scales that are \emph{regional}. For the horn and cone, the length of the shortest non-trivial closed curve through a given point (i.e. the circumference) is a regional property, and it is this length that the field modes are sensitive to.  Regional properties are those that cannot be inferred from the local geometry alone, yet do not depend on the global properties of the full manifold.

% The volume of the brane provides an energy due to its tension, however this volume is only sensitive to the length of the non-trivial geodesic 

In light of the physical relevance of a brane's position in the bulk, it is imperative to find a mechanism that would determine its location. Furthermore, without any preferred position, the brane position would correspond to a massless scalar (or scalars) that generically couples to matter with gravitational strength; this would be at odds with experimental constraints and therefore its stabilization is vital. In this work we exploit various contributions to an effective potential for the brane location, and have classified them according to effects which are local/geometric and global (or regional, in this approximation).

The most substantial calculation to be performed is that of the Casimir energy due to bulk fields, a global quantity.  Using the Casimir effect to stabilize branes and other moduli is not a new idea (see e.g \cite{Garriga:2000jb, Ponton:2001hq, Nasri:2002rx, Greene:2007xu}), however to our knowledge, it has not been used to stabilize a single brane;  this is a possibility if the bulk manifold lacks translation invariance. Calculation of Casimir energies is subtle because it strongly depends on the bulk and boundary geometry, topology, dimensionality, field type and boundary conditions.  On dimensional grounds, the magnitude of the effect can be estimated, however, in order to get the overall sign of the energy a full calculation usually must be performed.  Having explicit analytic expressions for the field modes makes this task much more tractable.
%One may find hope in recognizing that the mode structure of bulk quantum fields is sensitive to boundary and topological conditions, which ultimately means the vacuum expectation value of the total field energy depends on them; this is the Casimir effect, a generalization of the original parallel conducting plates example \cite{Casimir:1948dh}.  Given the existence of bulk fields with particular boundary conditions on the brane, the total Casimir (or vacuum) energy will depend on the location of the brane.  More particularly, there will be locations in the manifold where the Casimir energy is minimized, thus providing a localizing mechanism for the brane  -- the force due to the gradient in the Casimir energy.
%\footnote{The existence of a local minimum in the energy is necessary, but not sufficient, to stabilize the brane.  There must also be dissipation.  However, since the brane is, by assumption, coupled to a massless bulk field, dissipation will be a generic consequence of the acceleration caused by the Casimir force.}

\begin{figure}
  \begin{center}
    \includegraphics[scale=.7]{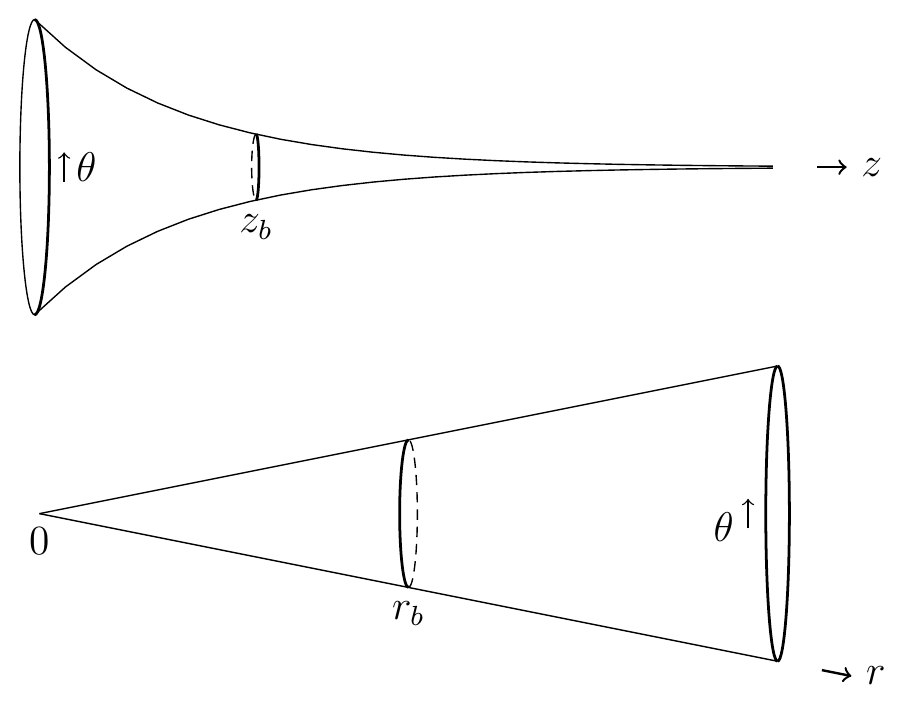}
  \end{center}
%\center{
%\beginpgfgraphicnamed{horn} 
%\begin{tikzpicture}[domain=0:5,scale=1.5]
% \draw[very thin,color=gray] (-0.1,-1.1) grid (3.9,3.9);
%\draw[->,line width=.7] (5.3,0) -- (5.5,0) node[right] {$z$};
%\draw[color=black] plot (\x,{exp(-\x)}) ;%node[right] {$f(x) = \frac{1}{20} \mathrm e^x$};
%\draw[color=black] plot (\x,{-exp(-\x)}) ;
%\draw (0,0) ellipse (.1cm and 1cm);
%\draw[->] (.2,-.1 ) -- (.2,.1);
%\node at (.35,0) {$\th$};
%\draw[thick] (0,-1) arc (-90:90: .1 and 1);
%\draw[thick] (2,-.135335)node[below] {$z_b$} arc (-90:90: .02 and .135335);
%\draw[densely dashed] (2,.135335)arc (90:270: .02 and .135335);
%\end{tikzpicture}
%\endpgfgraphicnamed 
%}
\caption{A partial embedding diagram of the horn and cone. The codimension-1 brane is pictured with coordinate $z_b$ and $r_b$, respectively.} \label{fig1}
\end{figure}

Here, we use the 2D hyperbolic horn and Euclidean cone, obtained from Euclidean 2-space by excising a wedge and identifying opposite sides of the cut  (giving a conifold, not a manifold), as models of an extra-dimensional manifold. Both have important (and complementary) features of more generic manifolds: one has intrinsic curvature and the other doesn't; one is infinite in extent while the other ends at a finite distance; they both have ``large" and ``small" regions and a corresponding breaking of translation invariance; and most importantly, scalar field modes can be solved for analytically on both.  They are interesting manifolds in their own right, but one may also consider them as approximations to a region of more complicated manifolds. %, where the zero-circumference point is at infinity.)  Perhaps more importantly, a brane wrapped around the horn can have zero extrinsic curvature; not so for the cone. The cone offers a complementary example as it has zero curvature and it is a finite distance to the vertex, unlike the horn.
%, i.e. the full spacetime manifold is  ${\cal M}^{4}\times {\cal H}^{2}/\Gamma$.

%, and a brane wrapped in a circle encompassing the cone's vertex, both brane properties and bulk-mode properties depend on the brane's distance from the vertex.

\begin{figure}
  \begin{center}
    \includegraphics[scale=.45]{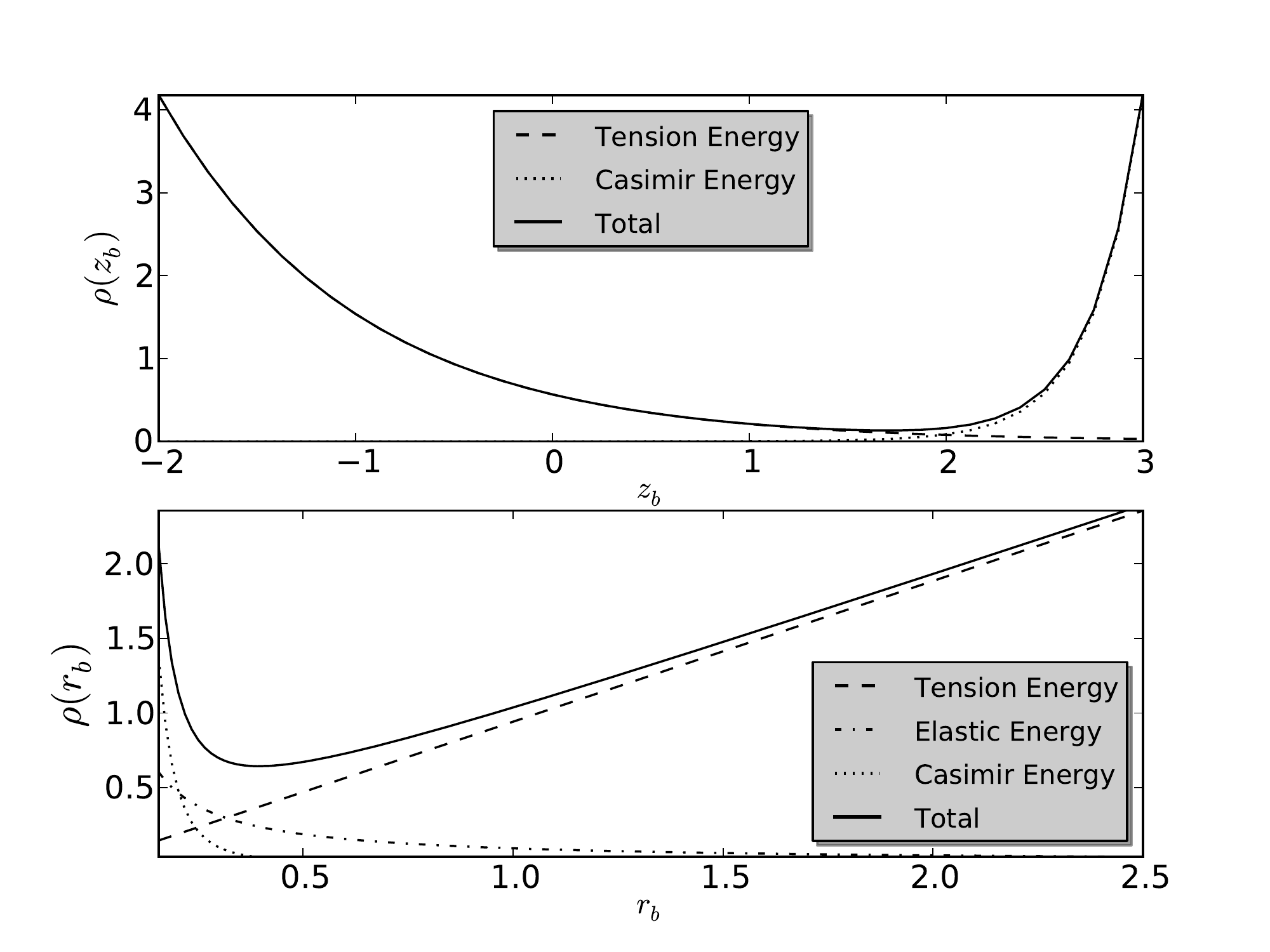}
  \end{center}
\caption{The total 4D energy density as a function of the brane position, for the horn and cone (top and bottom).}\label{fig2} %For illustration, we have taken the brane tension, $\s =1$ [$z_\star^{-5}$].
\end{figure}

We suppose that all Standard Model fields are confined to a codimension-1 brane (a 4-brane) as pictured in Figure \ref{fig1} and \ref{fig2}, but they may propagate freely in the compact dimension (i.e. it is ``universal" \cite{Appelquist:2000nn}).  For simplicity, we take a bulk scalar field, $\phi$, to exist and satisfy Dirichlet boundary conditions on the brane\footnote{This boundary condition might effectively arise from local interactions between $\phi$ and the localized matter/fields.}. We calculate the total energy, which is the sum of the local contributions due to geometric properties of the brane as well as that due to the zero point energy in $\phi$. Of course the geometry of the horn or the cone must still be stabilized in some way, but we do not consider this issue here.

The line element for the horn space-time can be written in coordinates such that
\begin{equation}
ds^2 = \e_{\m\n}dx^\m dx^\n  +e^{-2z/z_\star} z_\star^2 d\th^2 + dz^2 
\end{equation}
where $\eta_{\m\n}$ is the 4D Minkowski metric, we identify $(\th) \leftrightarrow (\th +2\p)$ and we have chosen our coordinates such that the horn circumference at $z=0$ is $2\p z_\star$ (from here on we shall work in units where the horn length scale, $z_\star \equiv 1$).  The geometry of the coned spacetime is flat and parameterized by a deficit angle, $2\p\de$, i.e.
%\begin{equation}
%ds^2 = \e_{\m\n}dx^\m dx^\n  + r^2 d\th^2 + dr^2 
%\end{equation}
identifying $\th \leftrightarrow \th +2\p\(1-\de\)$.  We will display salient results for the horn and summarize the results for the cone, saving many of the details and alternative configurations for a followup work \cite{Jacobs}.
\\
\emph{Local/Geometric Contributions}: Generally, there will be a non-zero brane tension, $\s$, that contributes an energy on the horn given by
\begin{equation}
E^{~}_{\text{ten}}=\int d^{4}x \sqrt{\abs{\g}} \s = 2\p e^{-z_b} V_{M}\s
\end{equation}
where $\g_{\m\n}$ is the induced metric on the 4-brane, and the volume of the regulated Minkowski spatial slice is denoted by $V_{M}$. There can also be an energy contribution due to the extrinsic curvature of the brane, $K_{ab}$, a type of elastic energy associated with how the brane bends within the bulk.  An example contribution would behave like $\int dV K^2$, where $K=K_{aa}$.  On this geometry, however $K$ is a constant, as are any contractions one could construct between $K_{ab}$ and the Riemann tensor of the bulk.  As a result any bending energy scales in the same way as that due to $\s$, and is thus indistinguishable. So without loss of generality we encode all geometric effects in $\s$. We note, however, the same is not true on the cone.
\\
\emph{Global Contributions}: 
%There will also be a vacuum energy associated with the standard model fields living on the brane.  Though we will not calculate their effect explicitly, by dimensional considerations, this energy must come in the following way
%\begin{equation}
%E_\text{0, SM} = \k_{SM} e^{(1+m)z_b} V_M
%\end{equation}
%where $\k_{SM}$ can be calculated in principle once $z_\star$ and $z_b$ have been specified/determined.
The modes of a massless bulk Klein-Gordon field, $\phi$, are (up to a normalization)
\begin{align}
u_{\bf i}&= e^{- i (\o t - {\bf p}\cdot {\bf x} - n\th )} Z_{n,k}(z)
%&= e^{- i (\o t - {\bf p}\cdot {\bf x} - \n\th )} R_{n,k}(r)~~~~~~~~~\text{(Cone)}
\end{align}
where ${\bf p}$ is the Minkowski momentum, $n\in \mathbb{Z}$ and %$\n\equiv n/(1-\de)$,
\begin{align}
%Z_{n,k}&=e^{z/2}\[I_{-i k}\(\abs{n} e^{z_L}\) I_{ik}\(\abs{n}  e^{z}\)- (k\leftrightarrow -k)\]\\
Z_{n,k}&\!=\!e^{z/2}\[a I_{ik}\(\abs{n}  e^{z}\)+b I_{-ik}\(\abs{n}  e^{z}\)\]& &(z_b \leq z)&\\
&\!=\!e^{z/2} K_{ik}\(\abs{n} e^{z}\)& &(z_b \geq z)&
\end{align}
where $I$ and $K$ are the modified Bessel functions of the first and second kind, respectively, and $a, b$ and the $\{k\}$ (which are implicitly dependent on $n$) are determined by normalization and the Dirichlet boundary condition, $Z_{n,k}(z_b)=0$. To make the problem more tractable, we regulate the infinite spatial volume of the horn by truncating the space at $z_L \ll z_b$, and impose there a Dirichlet boundary condition, taking $z_L\to -\infty$ in the end.
% We find that $u_{\bf i}=A_{\bf i} e^{- i (\o t - {\bf p}\cdot {\bf x} - n\th )} Z_{n,k}(z)$, where $n \in \mathbb{Z}$, $A_{\bf i}$ is a normalization constant, and $Z_{n,k}(z)$ satisfies the equation
%\begin{equation}
%Z_{n,k}'' - Z_{n,k}' + \(k^2 +\f{1}{4} - n^2 e^{2z}\)Z_{n,k}=0.
%\end{equation}
 %Note that while $\o$ doesn't have an explicit $n$-dependence, the $\{k\}$ do depend on $n$ implictly and are determined by the condition $Z_{n,k}(z_b)=0$. %The presence of constant negative curvature gives the spectrum a mass gap equal to $1/2$ in units of $z_\star^{-1}$.
%
%\begin{equation}\label{u_i, 2} 
%u_{\bf i}= T(t) X({\bf x})Z(z) \Theta(\th)
%\end{equation}
%from which one finds
%\begin{align}
%\ddot{T}(t)&=-\o^2 T(t)\\
%\del_x^2 X({\bf x})&= - {\bf p}^2 X({\bf x})\\
%\f{1}{L^2} \Theta''(\th)&=- Q^2 \Theta(\th)\\
%e^{-2z}\(  \f{Z''(z) - Z'(z)}{Z(z)}  + \o^2-{\bf p}^2\)&=Q^2
%\end{align}
 %The $d=0$ solutions may be obtained by simply setting ${\bf p}=0$.
%\begin{equation}
%Q= n/L, ~n \in \mathbb{Z}
%\end{equation}
%The function $Z_{n,k}(z)$ satisfies
%\begin{equation}
%Z_{n,k}''(z) - Z_{n,k}'(z) + \(k^2 +\f{1}{4} - e^{2z}\(\f{n}{L}\)^2\)Z_{n,k}(z)=0
%%F_{zz} -2F_{z} + (k^2+1)F - e^{2z}Q^2 F=0
%\end{equation}
We find that the $n=0$ modes do not contribute to the Casimir energy and so do not display them here. For $n \neq 0$, the positive frequency dispersion relation is $\o =  \sqrt{p^2 +k^2 +\f{1}{4}}$.%, where $p=\abs{{\bf p}}$ is the magnitude of momentum in the Minkowski directions.
% For the $n\neq0$ modes we find for $z_L \leq z \leq z_b$
%%\begin{align}\label{Z_soln_2}
%%Z_{n,k}&=e^{z/2}\[I_{-i k}\(q e^{z_L}\) I_{ik}\(q e^{z}\) -I_{ik}\(q e^{z_L}\) I_{-ik}\(q e^{z}\)\]\notag\\
%%&~~~~~~~~~~~~~~~~~~~~~~~~~~~~~~~~~~~~~~~~~~~~~~~~(z_L \leq z \leq z_b)\notag\\
%%&=e^{z/2} K_{ik}\(q e^{z}\)\notag\\
%%&~~~~~~~~~~~~~~~~~~~~~~~~~~~~~~~~~~~~~~~~~~~~~~~~(z_b \leq z < \infty)
%%\end{align}
%\begin{align}
%Z_{n,k}=e^{z/2}&\[I_{-i k}\(\abs{n} e^{z_L}\) I_{ik}\(\abs{n}  e^{z}\)\right. \\
%&-\left.I_{ik}\(\abs{n}  e^{z_L}\) I_{-ik}\(\abs{n}  e^{z}\)\]\notag
%\end{align}
%and for $z_b \leq z < \infty$
%\begin{equation}
%Z_{n,k}=e^{z/2} K_{ik}\(\abs{n} e^{z}\)
%\end{equation}
%\begin{align*}
%u &= \arctan x & dv &= 1 \, dx
%\\ du &= \frac{1}{1 + x^2} dx & v &= x.
%\end{align*}
%, and $ q=\abs{n}/L$.

We take a global approach to the Casimir effect, calculating the vacuum energy in $\phi$, for which we use the canonical result $E_0=\f{1}{2}\sum_{\bf i} \o_{\bf i}$. The sum, which is over the modes from both sides of the boundary, is clearly infinite and so we employ the zeta-function regularization technique (see e.g. \cite{Bordag:2009zz}):%\footnote{We work in units where $\hbar=c=1$}
\begin{equation}\label{zeta_fn_reg_E_0}
E_0=\lim_{s\to0}E_0(s)=\lim_{s\to0}\f{\m^{2s}}{2}\sum_{\bf i} \o_{\bf i}^{1-2s}%\equiv  \f{\m^{2s}}{2}\zeta_P(s-1/2)\\
\end{equation}  
%{\bf $\zeta_P(s-1/2)$ 
where $\m$, the renormalization scale, has units of energy.

%For ease of normalization we compactify the Minkowski dimensions using a torus of fundamental side length, $L_M$, which is taken to infinity at the end. In this limit the momenta in these directions become continuous and the sums are converted into integrals.
Because the summand is even in $n$, we may sum $n$ over twice the positive integers, obtaining from  \eqref{zeta_fn_reg_E_0} the expression
\begin{align}
E_0\(s\)%&= \f{\m^{2s} V_M}{2\p^2}\sum_{n=1}^{\infty}\sum_{\{k\}}\int dp p^2 \(p^2 + k^2 +\f{1}{4}\)^{1/2-s}\notag\\
\!\!= \f{\m^{2s} V_M}{8\p^{3/2}}\f{\Gamma(s-2)}{\Gamma(s-1/2)}\sum_{n=1}^{\infty}\sum_{\{k\}}   \( k^2 +\f{1}{4}\)^{2-s}
\end{align}
where we have used a Mellin transform to compute the $p$-integral. The $\{k\}$ are not known explicitly as they correspond to roots of Bessel functions, however the sum over the spectrum may be represented as a contour integral \cite{Bordag:2009zz}.
%\begin{align}\label{contour_rep}
%\sum_{\{k\}}\(\!k^2 \!+\! \f{1}{4}\!\)^{\!2-s} \!\!\!\!=\!\f{1}{2\p i}\oint_\g dk \!\(\!k^2 \!+\! \f{1}{4}\!\)^{2-s}\!\f{\d}{\d k}\! \ln\! \Delta_n(k)
%\end{align}
%where $\g$ is a counter-clockwise contour that encloses the entire spectrum (the positive real axis) and $\Delta_n(k)$ are mode-generating functions whose roots correspond to the spectrum of $k$ (see e.g.  \cite{Bordag:2009zz} or \cite{Fursaev:2011zz}).
%Because of our choice of Dirichlet conditions, they will be easily built out of the $k$-dependent part of our solutions, namely the $Z_{n,k}(z)$ 
With some mild restrictions on the choice of the generating functions used in that technique, one can show that contour may be deformed and, after analytic continuation in $s$ we arrive at
\begin{equation}\label{E_0_2}
E_0\!\(s\) \!=\! -\f{\m^{2s} V_M}{32\p^{2}} \! \sum_{n=1}^\infty \int_{1/2}^\infty \!\!\!\!dk   \( \!k^2 \!-\!\f{1}{4}\!\)^{2-s} \!\!\f{\d}{\d k}\! \ln \De_n(ik)
\end{equation} 
where an ${\cal O}(s)$ correction has been omitted for brevity and the $\Delta_n(k)$ are mode generating functions, found for both sides of $z_b$ to combine to effectively become
\begin{equation}%\ln{\[I_{k}\(q e^{z_b}\)K_{k}\(q e^{z_b}\)\]}
\ln{\Delta_n(i k)}\to\ln{\[I_{k}\(\abs{n} e^{z_b}\)K_{k}\(\abs{n} e^{z_b}\)\]} 
\end{equation}
%\begin{align}\label{rotated}
%\sum_{\{k\}}\(k^2 + \f{1}{4}\)^{2-s}=\f{\sin{\p s}}{\p}&\int_{1/2}^{\infty} dk \(k^2 - \f{1}{4}\)^{2-s}\notag\\\
%&\times \f{\d}{\d k} \ln \Delta_n(i k)%~~~\f{\sin{[\f{\p}{2} \(m+1-2s\) ]}}{1}
%\end{align}

%, noting that $\Gamma(s-1/2) = -2 \sqrt{\p}+{\cal O}(s)$ and $\Gamma(s-2)\sin{\p s} =  \p/2 + {\cal O}(s)$,

The divergent parts from $E_0(s)$ need to be isolated so that they may be either analytically continued to finite quantities or explicitly absorbed through some renormalization(s). We cannot yet perform an analytic continuation in $s$ of \eqref{E_0_2}, so we approach it using a uniform asymptotic expansion, closely following \cite{Bordag:2009zz}.
The divergences occur at large $k$ and $n$, and the expansion isolates the asymptotic behavior, taking these variables to infinity simultaneously while keeping their ratio fixed. 
The divergent behavior can then be understood analytically and properly dealt with.

To this end, we will decompose $E_0(s)$ into a sum of its divergent and finite parts %($s$-independent) It is then straightforward to analytically continue or explicitly absorb the divergent contributions through renormalization. 
\begin{equation}\label{sigma}
E_0(s) \equiv E_0^{\text{div}}(s)+E_0^{\text{fin}}
\end{equation}
We define $ x_b \equiv \abs{n} e^{z_b}$, $y \equiv k/x_b$, and perform a uniform asymptotic expansion of the modified Bessel functions (see e.g. \cite{abramowitz+stegun}) which results in the expansion
\begin{align}\label{KI_expansion}
\ln \Delta_n(i y x_b) \sim  &\ln{\[\f{1}{2x_b \sqrt{y^2+1}}\]} +\sum_{j=1}^\infty \f{{\cal F}_{2j}(y)}{(y x_b)^{2j}}
\end{align}
where the ${\cal F}_{2j}(y)$ are fractions of polynomials in $y$. Under these variable changes, we expand \eqref{E_0_2} in powers of $x_b$ which results in (after a bit of effort)
% \begin{align}\label{E0_of_y_and_xb}
%E_0\(s\) = -\f{\m^{2s} V_M}{32\p^{2}}  \sum_{n=1}^\infty  x_b^{4-2s}&\int_{(2 x_b)^{-1}}^{\infty} dy \(y^2 - \f{1}{4 x_b^{2}}\)^{2-s}\notag\\
%&\times\f{\d}{\d y} \ln \Delta_n(i y x_b)
%\end{align}
\begin{align}
E_0^{\text{div}}(s)=&\f{\m^{2s} V_M}{64\p^{2}}\sum_{n=1}^\infty{\bigg [}\f{12+29s}{192s}x_b^{-2s}\notag\\
&~~~+\f{-288+768s}{192s}x_b^{2-2s}+\f{1}{s}x_b^{4-2s}{\bigg]},%+ {\cal O}\(s\){\bigg]}
\end{align}
%\begin{align}
%E_0^{\text{div}}(s)\!=\!\f{\m^{2s} V_M}{64\p^{2}}\sum_{n=1}^\infty{\bigg [}\f{12+29s}{192s}x_b^{-2s}-(288-768s)x_b^{2-2s}+\f{1}{s}x_b^{4-2s}{\bigg]},%+ {\cal O}\(s\){\bigg]}
%\end{align}
which we subtract from $E_0(s)$ to (numerically) obtain $E_0^{\text{fin}}$. There is a  pole in $E_0^{\text{div}}(s)$ that does not depend on $z_b$, and so is simply discarded.

There will also be a vacuum energy associated with the standard model fields living on the brane.  Though we will not calculate their effect explicitly, by dimensional considerations this energy must behave approximately as $E_{0,\text{SM}} \sim \k_{SM} e^{4 z_b} V_M$. In principle $\k_{SM}$ can be calculated once $z_\star$ and $z_b$ have been specified/determined.\\
\emph{Summary of Results}: To summarize, the total energy is given by
\begin{align}
E_{\text{tot}}%&=E_0(s) + E_{\text{ten}}\notag\\
= E_{\text{ten}} + E_0^{\text{div}}(s) + E_0^{\text{fin}} + E_{0,\text{SM}} 
\end{align}
%\ssec{Asymptotics and Numerics}
%The total energy must be computed numerically, however w
We obtain analytic expressions for the asymptotic behavior,  providing a useful check of the numerics.  Each contribution to the (4D) energy density may be approximated using a  series or an Euler-Maclaurin expansion, giving% Asymptotically, we find %\footnote{Technically, this divergence is accounted for by appropriate geometric counter-terms in the effective action.}
\begin{align}
%\lim_{z_b\to -\infty}\f{E}{V_M} &\sim \( \f{-1.5 \times 10^{-4}}{z_\star^5}+2\p\s\)L e^{-z_b/z_\star}\\  
\lim_{z_b\to -\infty}\r(z_b) &\sim 2\p\s_\text{ren} z_\star e^{-z_b/z_\star}\\%-1.5 \times 10^{-4}  z_\star^{-5}
%\notag\\
\lim_{z_b\to \infty}\r(z_b) &\sim\( 2.5\times10^{-5}+\k_\text{SM}\)\times\f{e^{4z_b/z_\star}}{z_\star^4}
%\lim_{z_b\to \infty}\f{E}{V_M}\sim &-\f{3}{64\p^2}\zeta'(-2)\f{e^{2z_b/z_\star}}{L^2}+\f{1}{32\p^2}\zeta'(-4)\f{e^{4z_b/z_\star}}{L^4}
\end{align}
where we have put back the horn curvature scale, $z_\star$, and have renormalized the brane tension to include finite quantum corrections from the bulk scalar. For $\s_\text{ren} >0$, a local minimum of the effective potential occurs, assuming the sign and magnitude of $\k_\text{SM}$ doesn't spoil the vacuum effect of $\phi$. For an ${\cal O}(1)$ brane tension (in units of $z_\star$), the effective mass for the position modulus is ${\cal O}(z_\star^{-1})$. In \cite{Jacobs} we will show qualitatively similar results for the cone.  A plot of the effective potential for the brane position modulus is plotted for both geometries in Figure \ref{fig2}.

%\section{Conclusions}\label{conclusions}
%As a mechanism for localizing and stabilizing a brane position, we have considered the application of the Casimir effect, in conjunction with a (classical) brane tension, using the spacetime ${\cal M}^{4}\times {\cal H}^{2}/\Gamma$ as a toy model.  This was achieved by using a bulk scalar with a Dirichlet brane condition. In the absence a brane tension, the total energy is unbounded from below, however for sufficiently large tensions, 

Interestingly, it is the competition between tension and Casimir energies (i.e. between a local and global effect) on the horn that results in a stable brane position, while on the cone this can be achieved even in the absence of Casimir, from the competition between just the geometric contributions. Since we have used a massless bulk field a residual scaling symmetry remains since there is only one dimensional parameter -- the spatial curvature (on the horn), and the distance to the vertex (on the cone).   Consequently the position dependence of each contribution to the total energy is monotonic, so the force resulting from each can only push the brane to positive or negative infinity on the horn, or zero and infinity on the cone.  We therefore see why the competition between at least two effects was needed.

% (on these two geometries) all of these effects behave monotonically with the brane position, the competition between at least two effects was needed.

%Again, this is the only dimensional quantity, so brane localization and stabilization must be a result of competition among forces with different position dependence. 

%However the energy due to two or more of these physical effects  can have a minimum at a finite position. In other words, multiple forces exist on such a brane, and their combination may localize and stabilize the brane.

More generic compact spaces have no residual continuous symmetries, and we can expect  each contribution to the total energy to have local extrema within the manifold. In this context, the Casimir energy of bulk fields with boundary conditions on the brane is particularly interesting for 3-branes in (3+d+1)-dimensional space
(a standard braneworld scenario) because it may be the only source of bulk-position-dependent energy density (and thus the only cause of localization and stabilization) when such branes are point-like in the extra dimensions.
%However, this won't be the case on a more complex manifold where there is even less symmetry; generally there should be many special points capable of providing energy extrema. Similarly, it should be unnecessary for the brane to be extended in the extra dimensions as no geometrical contributions to the energy density (which emerge only for non-point-like branes) should be necessary.

In a followup work \cite{Jacobs}, we will provide more details of this calculation, both in 2+1 and 5+1 dimensions, and also the results for a Neumann boundary condition.
%Other variants that would be illuminating to consider include using massive fields, fields of different spin, or variants of the manifold, e.g. embedding this in an expanding FRW background and/or using more compact dimensions; this last option allowing for the possible interplay amongst many scales.
The mechanism considered here is but one example of a generic class of effects wherein the physical behavior of higher-dimensional systems is sensitive to the structure of finite regions of the manifold.    This should prove to be of relevance to model builders with specific phenomenological goals in mind, and will not be limited to braneworld scenarios alone.\\
   
%{\bf (Dave: Comments on possible applications for degravitation? Brane bending mode frozen?--For comparison, does a brane tension alone not give the brane-bending mode a mass?)}

We would like to thank Ling-yi Huang, Harsh Mathur and Claudia de Rham for many useful discussions. This work is supported by grants from the U.S. Department of Energy and Department of Education.

\bibliographystyle{unsrt}
\bibliography{Casimir_bib}
%\renewcommand{\bibfont}{\small}
%\begin{multicols}{2}[\printbibheading]
%\printbibliography[heading=none]
%\end{multicols}

%\begin{thebibliography}{99}
%
%\bibitem{ArkaniHamed:1998rs} 
%  N.~Arkani-Hamed, S.~Dimopoulos and G.~R.~Dvali,
%  %``The Hierarchy problem and new dimensions at a millimeter,''
%  Phys.\ Lett.\ B {\bf 429}, 263 (1998)
%  [hep-ph/9803315].
%  %%CITATION = HEP-PH/9803315;%%
%
%
%\end{thebibliography}

\end{document}